\documentclass[aps,prc,twocolumn, showpacs,superscriptaddress,floatfix]{revtex4-1}  % for review and submission
\usepackage{graphicx}  % needed for figures
\usepackage{dcolumn}   % needed for some tables
\usepackage{bm}        % for math
\usepackage{amssymb}   % for math
\usepackage{xcolor}
\usepackage{hyperref}
\RequirePackage{lineno}

\graphicspath{{./Figures/}}

\newcommand{\sNN}[1]{$\sqrt{s_{\rm NN}} = #1$}
\newcommand{\pp}{$p+p$}
\newcommand{\vn}[1]{v_{2}\lbrace{\rm#1}\rbrace}
\newcommand{\vHFE}{v_2^{eHF}}
\newcommand{\HFE}{$e_{\rm HF}$}
\newcommand{\vPho}{v_2^{\rm pho}}
\newcommand {\pt}	{p_T}
\newcommand {\nse}	{n\sigma_{\rm electron}}
\newcommand {\effPho}	{\varepsilon_{\rm pho}}
\newcommand {\nPho}	{N_{\rm pho}}
\newcommand {\nHFE}	{N_{\rm eHF}}
\newcommand {\Ke}	{K_{\rm e3}}
\newcommand {\psiEP}	{\Psi_{\rm EP}}

\newcommand {\psiRP}	{\Psi_{\rm RP}}
\newcommand{\gevc}{\mbox{$\mathrm{GeV/}c$}}

\newcommand{\dedx}{\mbox{$dE/dx$}}

\begin{document}
	
\title{Elliptic flow of electrons from heavy-flavor hadron decays in Au+Au collisions at $\sqrt{s_{\rm NN}} = $ 200, 62.4, and 39 GeV}

\affiliation{AGH University of Science and Technology, FPACS, Cracow 30-059, Poland}
\affiliation{Argonne National Laboratory, Argonne, Illinois 60439}
\affiliation{Brookhaven National Laboratory, Upton, New York 11973}
\affiliation{University of California, Berkeley, California 94720}
\affiliation{University of California, Davis, California 95616}
\affiliation{University of California, Los Angeles, California 90095}
\affiliation{Central China Normal University, Wuhan, Hubei 430079}
\affiliation{University of Illinois at Chicago, Chicago, Illinois 60607}
\affiliation{Creighton University, Omaha, Nebraska 68178}
\affiliation{Czech Technical University in Prague, FNSPE, Prague, 115 19, Czech Republic}
\affiliation{Nuclear Physics Institute AS CR, 250 68 Prague, Czech Republic}
\affiliation{Frankfurt Institute for Advanced Studies FIAS, Frankfurt 60438, Germany}
\affiliation{Institute of Physics, Bhubaneswar 751005, India}
\affiliation{Indiana University, Bloomington, Indiana 47408}
\affiliation{Alikhanov Institute for Theoretical and Experimental Physics, Moscow 117218, Russia}
\affiliation{University of Jammu, Jammu 180001, India}
\affiliation{Joint Institute for Nuclear Research, Dubna, 141 980, Russia}
\affiliation{Kent State University, Kent, Ohio 44242}
\affiliation{University of Kentucky, Lexington, Kentucky, 40506-0055}
\affiliation{Lamar University, Physics Department, Beaumont, Texas 77710}
\affiliation{Institute of Modern Physics, Chinese Academy of Sciences, Lanzhou, Gansu 730000}
\affiliation{Lawrence Berkeley National Laboratory, Berkeley, California 94720}
\affiliation{Lehigh University, Bethlehem, PA, 18015}
\affiliation{Max-Planck-Institut fur Physik, Munich 80805, Germany}
\affiliation{Michigan State University, East Lansing, Michigan 48824}
\affiliation{National Research Nuclear University MEPhI, Moscow 115409, Russia}
\affiliation{National Institute of Science Education and Research, Bhubaneswar 751005, India}
\affiliation{National Cheng Kung University, Tainan 70101 }
\affiliation{Ohio State University, Columbus, Ohio 43210}
\affiliation{Institute of Nuclear Physics PAN, Cracow 31-342, Poland}
\affiliation{Panjab University, Chandigarh 160014, India}
\affiliation{Pennsylvania State University, University Park, Pennsylvania 16802}
\affiliation{Institute of High Energy Physics, Protvino 142281, Russia}
\affiliation{Purdue University, West Lafayette, Indiana 47907}
\affiliation{Pusan National University, Pusan 46241, Korea}
\affiliation{Rice University, Houston, Texas 77251}
\affiliation{University of Science and Technology of China, Hefei, Anhui 230026}
\affiliation{Shandong University, Jinan, Shandong 250100}
\affiliation{Shanghai Institute of Applied Physics, Chinese Academy of Sciences, Shanghai 201800}
\affiliation{State University Of New York, Stony Brook, New York 11794}
\affiliation{Temple University, Philadelphia, Pennsylvania 19122}
\affiliation{Texas A\&M University, College Station, Texas 77843}
\affiliation{University of Texas, Austin, Texas 78712}
\affiliation{University of Houston, Houston, Texas 77204}
\affiliation{Tsinghua University, Beijing 100084}
\affiliation{University of Tsukuba, Tsukuba, Ibaraki, Japan,}
\affiliation{Southern Connecticut State University, New Haven, Connecticut, 06515}
\affiliation{United States Naval Academy, Annapolis, Maryland, 21402}
\affiliation{Valparaiso University, Valparaiso, Indiana 46383}
\affiliation{Variable Energy Cyclotron Centre, Kolkata 700064, India}
\affiliation{Warsaw University of Technology, Warsaw 00-661, Poland}
\affiliation{Wayne State University, Detroit, Michigan 48201}
\affiliation{World Laboratory for Cosmology and Particle Physics (WLCAPP), Cairo 11571, Egypt}
\affiliation{Yale University, New Haven, Connecticut 06520}

\author{L.~Adamczyk}\affiliation{AGH University of Science and Technology, FPACS, Cracow 30-059, Poland}
\author{J.~K.~Adkins}\affiliation{University of Kentucky, Lexington, Kentucky, 40506-0055}
\author{G.~Agakishiev}\affiliation{Joint Institute for Nuclear Research, Dubna, 141 980, Russia}
\author{M.~M.~Aggarwal}\affiliation{Panjab University, Chandigarh 160014, India}
\author{Z.~Ahammed}\affiliation{Variable Energy Cyclotron Centre, Kolkata 700064, India}
\author{N.~N.~Ajitanand}\affiliation{State University Of New York, Stony Brook, New York 11794}
\author{I.~Alekseev}\affiliation{Alikhanov Institute for Theoretical and Experimental Physics, Moscow 117218, Russia}\affiliation{National Research Nuclear University MEPhI, Moscow 115409, Russia}
\author{D.~M.~Anderson}\affiliation{Texas A\&M University, College Station, Texas 77843}
\author{R.~Aoyama}\affiliation{University of Tsukuba, Tsukuba, Ibaraki, Japan,}
\author{A.~Aparin}\affiliation{Joint Institute for Nuclear Research, Dubna, 141 980, Russia}
% ...

\collaboration{STAR Collaboration}\noaffiliation


\begin{abstract}
We present measurements of elliptic flow ($v_2$) of electrons from the decays of heavy-flavor hadrons (\HFE) by the STAR experiment. For Au+Au collisions at $\sqrt{s_{\rm NN}} = $ 200 GeV we report $v_2$, for transverse momentum ($\pt$) between 0.2 and 7~\gevc, using three methods: the event plane method ($\vn{EP}$), two-particle correlations ($\vn{2}$), and four-particle correlations ($\vn{4}$). For Au+Au collisions at $\sqrt{s_{\rm NN}}$ = 62.4 and 39 GeV we report $\vn{2}$ for $\pt< 2 \, \gevc$. $\vn{2}$ and $\vn{4}$ are non-zero at low and intermediate $\pt$ at 200 GeV, and $\vn{2}$ is consistent with zero at low $\pt$ at other energies.The $\vn{2}$ at the two lower beam energies is systematically lower than at $\sqrt{s_{\rm NN}} = $ 200 GeV for $\pt < 1 \  \gevc$. This difference may suggest that charm quarks interact less strongly with the surrounding nuclear matter at those two lower energies compared to \sNN{200}~GeV.

\end{abstract}

\maketitle

\section{\label{sec:introduction}Introduction}
Experiments of ultrarelativistic heavy-ion collisions aim to
create deconfined strongly-interacting matter, a Quark-Gluon Plasma (QGP),
and to study the QGP properties~\cite{STAR:white:paper,PHENIX:white:paper,PHOBOS:white:paper,BRAHMS:white:paper}. Heavy quarks (charm and bottom) provide a unique probe of the QGP properties~\cite{Andronic:2015wma,Averbeck:2013oga, Prino:2016cni}:
because their masses are large compared with the thermal energy 
expected in heavy-ion collisions~\cite{Rapp:HF:review},
they are mainly produced in interactions with high momentum transfer, 
very early in the heavy-ion collisions and they are expected to interact 
with the QGP differently than light and strange quarks~\cite{Dokshitzer:2001zm, Armesto:2003jh, Djordjevic:2005db, WHDM:2005gt}. For example, the Djordjevic-Gyulassy-Levai-Vitev (DGLV)~\cite{WHDM:2005gt} theory successfully describes the observed light hadron quenching with gluon radiation alone, while additional collisional energy loss is required for charm and bottom quarks. Moreover, heavy quark production is sensitive to the dynamics of the nuclear medium
created in the collisions~\cite{HQ:thermalization};  measurements of their production and elliptic flow $v_2$ could be 
used to determine the fundamental properties of the QGP, such as transport coefficients (see, for instance, Ref.~\cite{HQ:transport} and references therein). Electrons from the decays of heavy flavor hadrons (\HFE) 
represent well the directions of the parent D (B) mesons when the transverse momentum ($\pt$) of the electron is $p_T > 1.5 (3)$ \gevc~\cite{WenqinXu:PhD,WenqinXu:Panic2011}. Thus \HFE\ $v_2$ serves as a good proxy for heavy quark $v_2$, particularly at high transverse momenta. At lower $\pt$ \HFE\ still carries information about the parent meson $v_2$, even though it is diluted by the decay kinematics~\cite{Batsouli:2002qf}.

Heavy quark in-medium interactions have been studied both at the Relativistic Heavy Ion Collider (RHIC) and the Large Hadron Collider (LHC). Energy loss is experimentally investigated by the nuclear modification factor $R_{AA}$, which is defined as the yield in heavy-ion collisions divided by that in p+p scaled by the number of binary collisions. Both the STAR and PHENIX experiments reported a strong suppression of \HFE\ production at high transverse momenta at mid-rapidity in central Au+Au collisions at $\sqrt{s_{\rm NN}} = $ 200 GeV~\cite{Adare:2006nq,Abelev:2006db,Adare:2010de}, relative to \HFE\ produced in p+p collisions. No significant attenuation of the \HFE\ yield was observed in d+Au collisions~\cite{Abelev:2006db,Adare:2012yxa}. Moreover, the charmed meson $R_{AA}$ (measured via the full reconstruction of hadronic decay of $D^0$) in central Au+Au collisions at that energy~\cite{Adamczyk:2014uip} shows a strong suppression for $\pt>3 \, \gevc$. These results indicate that heavy quarks lose energy while traversing a dense strongly interacting medium created in heavy-ion collisions. The LHC experiments observed a similar situation in heavy-ion collisions at $\sqrt{s_{\rm NN}} = $ 2.76 TeV: heavy flavor production (studied either via charmed mesons~\cite{Adam:2015nna,ALICE:2012ab}, semi-leptonic decays of heavy flavor hadrons at forward rapidity~\cite{Abelev:2012qh}, $J/\psi$ from B-hadron decays~\cite{Chatrchyan:2012np} or b-flavored jets~\cite{Chatrchyan:2013exa}) is suppressed in central Pb+Pb collisions compared to the p+p case. Furthermore, a non-zero, positive elliptic flow of \HFE\ and $\mu_{\rm HF}$ was detected at the top RHIC~\cite{Adare:2006nq,Adare:2010de} energy and at the LHC~\cite{Adam:2015pga, ALICE:HF:v2} at low and intermediate $\pt$. Those data suggest a collective behavior of heavy quarks (mainly charm) with low transverse momenta. Charmed meson $v_2$ measured at the LHC~\cite{Abelev:2013lca} and RHIC~\cite{Lomnitz:2016rpz} supports this interpretation.

One of the difficulties in interpretation of the $v_2$ results is that various methods have different sensitivities to elliptic flow fluctuations and to particle correlations not related to the reaction plane, so-called non-flow. Jets and resonance decays are considered to be the most important sources of these non-flow correlations. In this paper, we present the STAR measurements of the \HFE\ $v_2$ using two- and four-particle correlations~\cite{Borghini:2000sa} 
($\vn{2}$ and $\vn{4}$, respectively) and the event plane method ($\vn{EP}$)~\cite{Poskanzer:1998yz} in Au+Au collisions at $\sqrt{s_{\rm NN}} = $ 200 GeV
at RHIC.  In the case of $\vn{2}$ and $\vn{EP}$, there are positive contributions from both $v_2$ fluctuations and non-flow (the event plane and two-particle correlation methods are approximately equivalent~\cite{Trainor:2008jp}). 
When $v_2$ is obtained with four-particle correlations ($\vn{4}$), the fluctuations give a negative contribution and non-flow is suppressed. Therefore, $\vn{2}$ gives an upper limit, and $\vn{4}$ gives a lower limit, on elliptic flow~\cite{Voloshin:2007pc}.

The heavy flavor nuclear modification factor and elliptic flow at the top RHIC energy indicate that heavy quarks interact strongly with the QGP. RHIC Beam Energy Scan results show that elliptic flow of inclusive charged hadrons is approximately independent of beam energy in the range of 39-62.4 GeV (the difference is less than 10\% for $0.5<\pt<3$~GeV/c)~\cite{STAR:BES:iclusive:hadron:v2}. Current data on the \HFE\ $R_{AA}$ and $v_2$ in Au+Au collisions at $\sqrt{s_{\rm NN}} = $ 62.4 GeV are inconclusive about whether heavy quarks interact with a nuclear medium at that lower energy as strongly as at $\sqrt{s_{\rm NN}} = $ 200 GeV. We present new measurements of the \HFE\ $\vn{2}$ in Au+Au collisions at $\sqrt{s_{\rm NN}} = $ 62.4 and 39 GeV. The \HFE\ $\vn{2}$ at these energies could provide information about the energy dependence of the strength of heavy quark interactions with a hot and dense nuclear medium.

\section{\label{sec:analysis}Data analysis}

Three main STAR subsystems are used in this analysis: the Time Projection Chamber (TPC)~\cite{tpc_det}, the Barrel Electromagnetic Calorimeter (BEMC)~\cite{bemc_det} and the Time-of-Flight (ToF)~\cite{tof_det} detectors. These detectors provide tracking and particle identification.

The data used in this analysis were obtained using minimum-bias and high-$\pt$ (so-called high tower~\cite{STAR:NPE:pp200GeV}) triggers. 
The minimum-bias trigger was defined as a coincidence signal in the east and west vertex position detectors (VPDs)~\cite{ref:vpd_det} located 5.7~m from the interaction point, in the pseudo-rapidity range of $4.2 \leq \eta \leq 5.1$. The high tower triggers required at least one BEMC tower passing a given transverse energy threshold. We used cascading triggers with thresholds of $\sim2.6$~GeV, $\sim3.5$~GeV and $\sim4.2$~GeV. Collision centrality is determined using the number of reconstructed tracks in the TPC within $|\eta|< 0.5$~\cite{STAR:pi:pTspectra:200GeV}. Events with primary vertices located within $\pm30$~cm of the TPC's geometrical center along the beam direction and with 0-60\% centrality are selected for the $v_2$ measurement.  
The data samples used in this study are summarized in Tab.~\ref{Tab:dataset}. The number of high tower events correspond to $6.34 \times 10^9$ minimum bias events within the analyzed centrality range.  

We select tracks with at least 20 points measured in the TPC and at least 52\% of the maximum number of possible TPC points (which is 45 at midrapidity) to remove split tracks (one track reconstructed as two or more in the TPC). The distance-of-closest-approach (DCA) in the three-dimensional space of a track to the collision vertex is required to be less than 1.5 cm, which corresponds to 3 standard deviations of the DCA distribution. 

Electrons are identified using the ionization energy loss ($\dedx$) in the TPC, the time-of-flight in the ToF detector and the energy deposited in BEMC towers. First, we select tracks with $|\eta|<0.7$ and $0<\nse<3$, where $\nse$ is the number of standard deviations from the expected mean $\dedx$ for electrons in the TPC. The $\nse$ cut was chosen to optimize the purity (to reduce a potential systematic error due to hadron contamination) and the available statistics (which is crucial for the $\vn{4}$ measurement). For $\pt<1$~\gevc, the velocity $\beta$ measured in the ToF is used to reject kaons: we require $|1-1/\beta|<0.03$ at 200 GeV, $-0.03< 1-1/\beta<0.02$ at 62.4 GeV and $-0.03< 1-1/\beta<0.01$ at 39 GeV. Different cuts are used because of the slightly different ToF resolution at different energies~\cite{STAR:VPD}. To further enhance electron identification at 39 and 62.4 GeV, we impose a more stringent requirement on $\nse$ ($0<\nse<2$) for these collision energies. In the $\pt$ range where the proton $\dedx$ band overlaps with the electron band ($1<\pt<1.5$~\gevc), we apply an additional cut of $|1-1/\beta|<0.1$ in order to reduce proton contamination. Finally, at $\pt>1 \ \gevc$, we select tracks that have a momentum-to-energy ratio in the range of $0.3 < pc/E < 2$, where $E$ is the energy of a single BEMC tower associated with a TPC track. The BEMC has a Shower Maximum Detector (SMD), which is a proportional gas chamber with strip readout at a depth of five radiation lengths designed to measure shower shapes and positions in the pseudorapidity - azimuthal angle ($\eta - \phi$) plane, and used to discriminate between electrons and hadrons. In order to further improve the purity of the electron sample, we require tracks to occupy more than one strip in both $\phi$ and $\eta$ SMD planes. 

\begin{table*}[htp]
%\begin{ruledtabular}
\begin{tabular}{lc}
\hline \hline 
Collision energy $\sqrt{s_{\rm NN}}$ & Data sample [million events] \\
\hline 
200 GeV (minimum bias trigger) & 142 \\
200 GeV (high tower trigger) & 41  \\ %(use $\mathcal{L}$ instead?) \\
62.4 GeV (minimum bias trigger) & 39  \\
39 GeV (minimum bias trigger) & 87  \\
\hline \hline 
\end{tabular}
%\end{ruledtabular}
\caption{\label{Tab:dataset} Au+Au data samples used for the analysis. The numbers represent $0-60\%$ most central events.}
\end{table*}

\begin{figure*}[htp]
\begin{center}
\begin{tabular}{cc}
\includegraphics[width=0.45\textwidth]{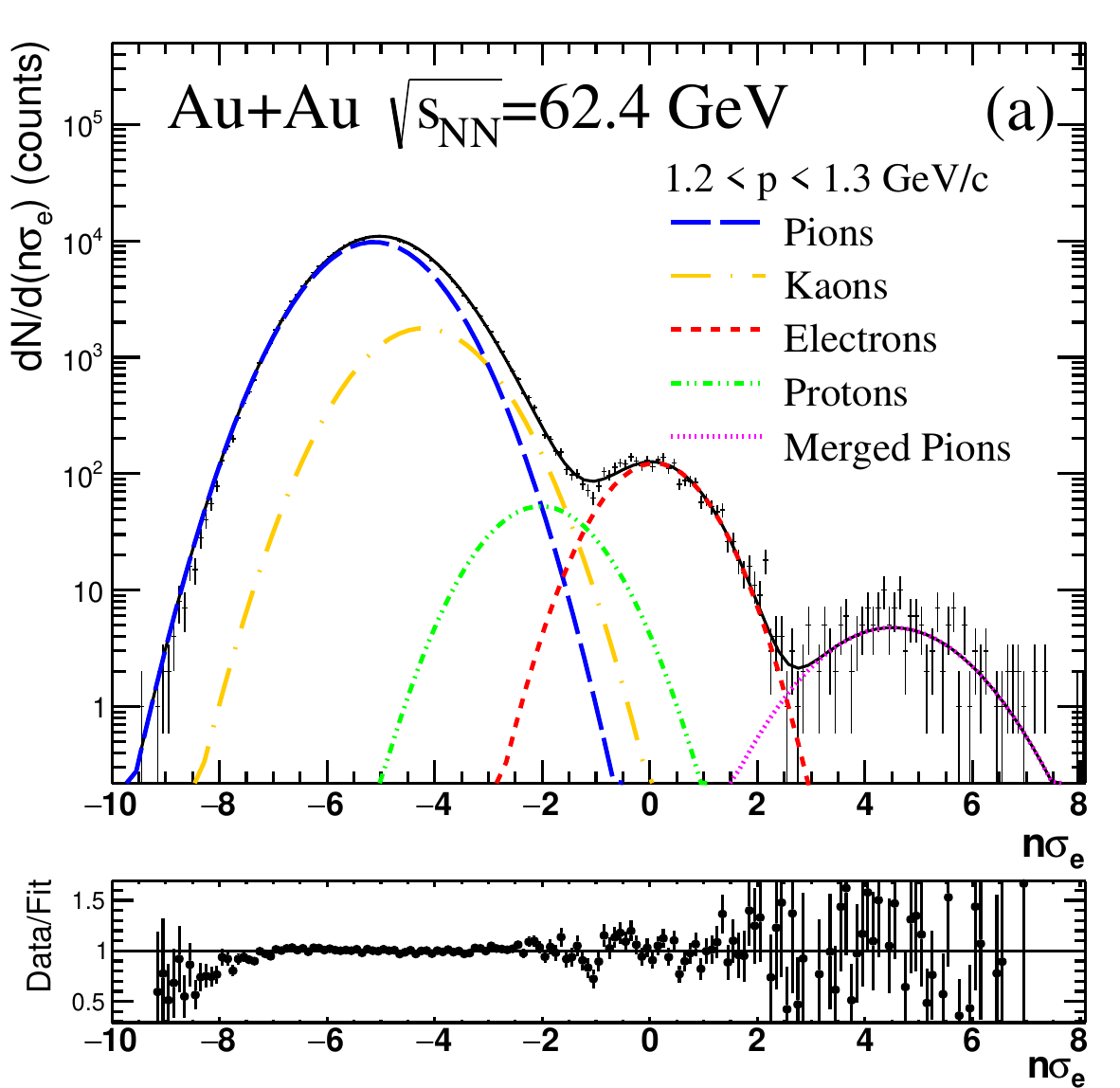} &
\includegraphics[width=0.45\textwidth]{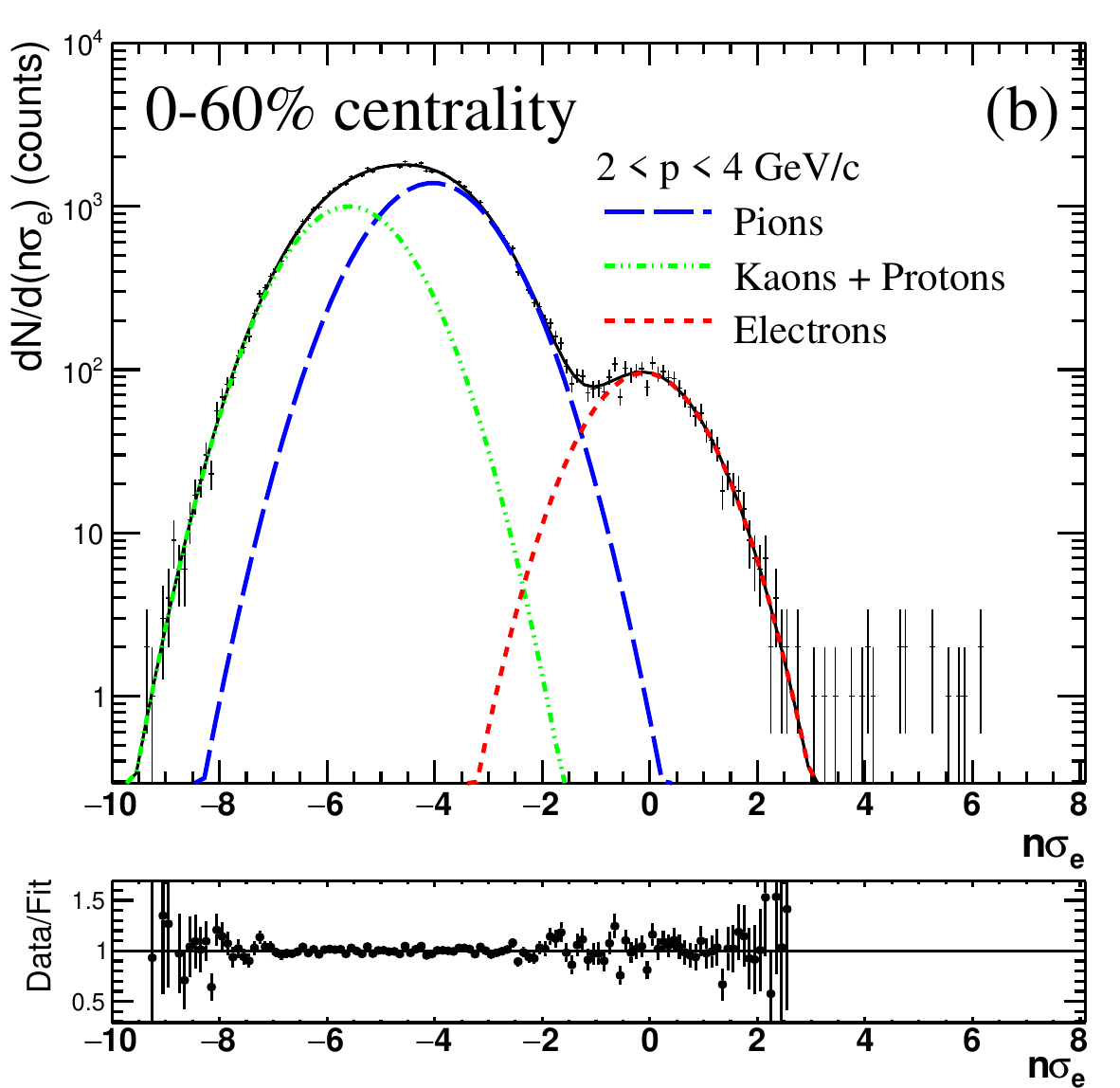} \\
\end{tabular}
\caption{\label{Fig:PurityFits} (Color online) Examples of $n\sigma_e$ distribution with fits for different hadronic components for minimum bias Au+Au collisions at \sNN{62.4}~GeV at low (a) and high momenta (b).} 
\end{center}
\end{figure*}

Hadron contamination is estimated by first fitting a sum of Gaussian functions for charged hadrons and electrons to the $\nse$ distribution in momentum bins, after applying all electron identification and track quality cuts, except the cut on $\nse$ itself.  Figure \ref{Fig:PurityFits} shows examples of such fits for the $0.9<p<1$~\gevc\ and $2<p<4$~\gevc\ bins for 62.4 GeV data. In Fig.~\ref{Fig:PurityFits}(a), we also include a Gaussian for merged pions that arise from track merging due to the finite two-track resolution of the TPC; these have a \dedx\ approximately two times larger than ``regular'' pions. Parameters of the Gaussian functions (mean and width) for each fit component are constrained using high-purity electron and hadron samples. The parameters for electrons are fixed based on an electron sample from photon conversion in the detector material and the Dalitz decay of $\pi^0$ and $\eta$ mesons. These electrons were identified by selecting  $e^+e^-$ pairs with a low invariant mass ($m_{e+e-}<0.15$~$\rm{\gevc}^2$); we describe this procedure in the next paragraph. 

\begin{figure}[htp]
\begin{center}
\begin{tabular}{c}
\includegraphics[width=0.45\textwidth]{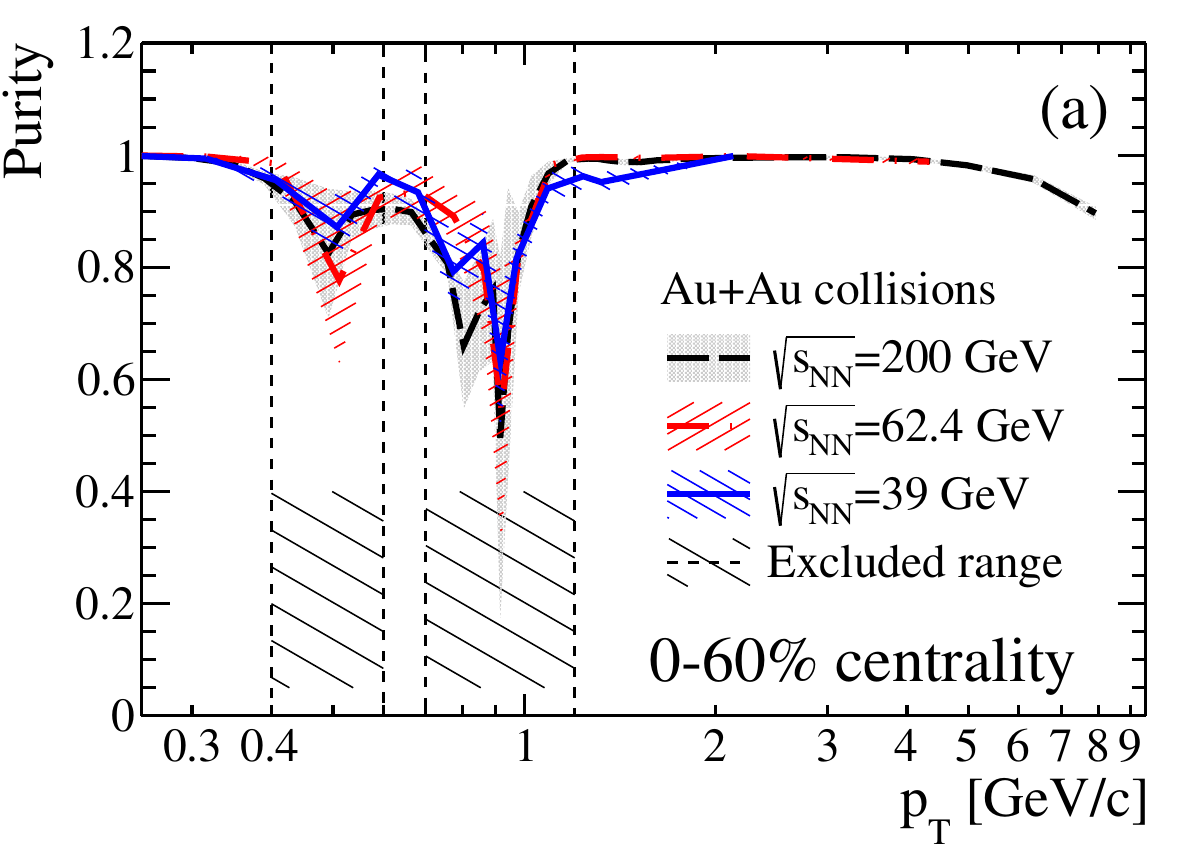} \\
\includegraphics[width=0.45\textwidth]{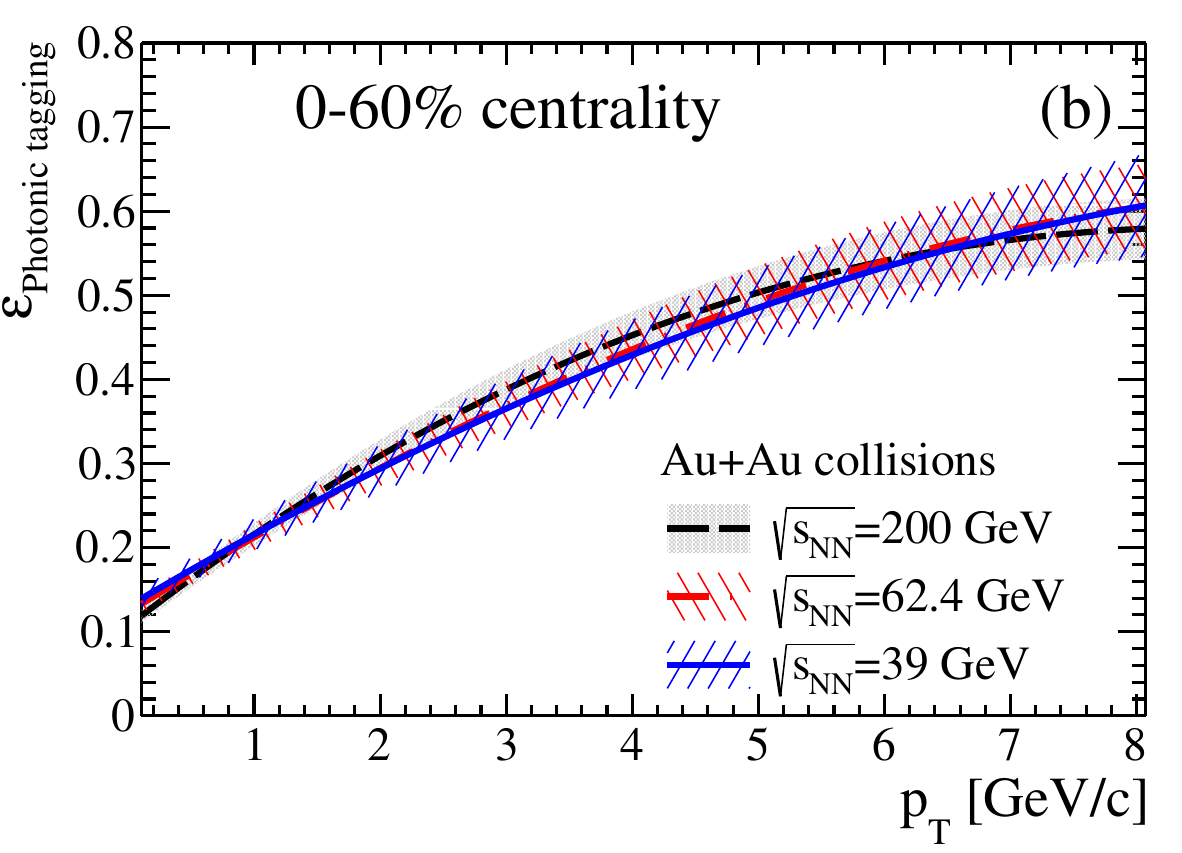} \\
\end{tabular}
\caption{\label{Fig:PurityPheRecoEff}(Color online) Electron purity (a) and photonic electron tagging efficiency (b). The bands show the combined systematic and statistical uncertainties. Centrality classes are indicated in the plot.} 

\end{center}
\end{figure}

For hadrons, we use the ToF at low and intermediate momenta to select tracks with a mass close to the mass expected for that specific hadron. At $p> 1.5$~\gevc, pions from $K_s^0$ decays are selected, which are identified via secondary vertex reconstruction. At high momenta a simplified fit model (three Gaussian functions: for electrons, pions and protons combined with kaons) describes the $\nse$ distribution well (see Fig. \ref{Fig:PurityFits}(b)). To improve fitting in the ranges where the kaon and the proton \dedx\ bands overlap with the electron band, we impose constraints on the hadron amplitudes: the amplitude of a Gaussian for a hadron is limited by the values determined outside of the crossing range, where hadron-electron separation is feasible. The Gaussian fits in $\nse$ bins are then used to calculate the hadron yields within the $\nse$ range selected for the analysis. Purity is defined as a ratio of electrons to all tracks that passed the quality and electron identification cuts. The width of the momentum bins is determined by the available statistics. At low $p$ we use narrow bins (widths of 50 or 100 MeV/c) and at higher momentum ($p>3\ \gevc$ for 200 GeV and $p>2\ \gevc$ for lower energies) we adopted bin widths of 1 or 2 GeV/c. The relativistic rise of pion $dE/dx$ within a wide momentum bin could lead to a non-Gaussian shape of the pion $\nse$ distribution. To quantify how much this affects our measurement, we compared the purity in the momentum range of $3 < p < 6 \ \gevc$ obtained with very narrow bins (50 MeV/c) with that using a wide bin of $3 < p < 6 \ \gevc$. As the results from these two choices of binning are consistent, the binning does not have a significant effect on the purity. The purity as a function of $\pt$ is finally calculated using a correlation between the inclusive electron $\pt$ and momentum, the uncertainty on which is included in the systematic uncertainty evaluation. Figure \ref{Fig:PurityPheRecoEff} (a) shows the purity as a function of $\pt$. The results have similar shapes for all data sets. The overall purity is 90\% or better and hadron contamination is only significant for $\pt \sim 0.5-0.6$~\gevc\ and $\pt \sim 0.8-1.1$~\gevc\ due to the overlap of the kaon and the proton \dedx\ bands. To minimize systematic uncertainty due to hadron contamination, we removed the $\pt$ bins of $0.5-0.6$~\gevc\ and $0.7-1.2$~\gevc\ from the analysis. 

\begin{figure}[htp]
\begin{center}
\begin{tabular}{c}
\includegraphics[width=0.45\textwidth]{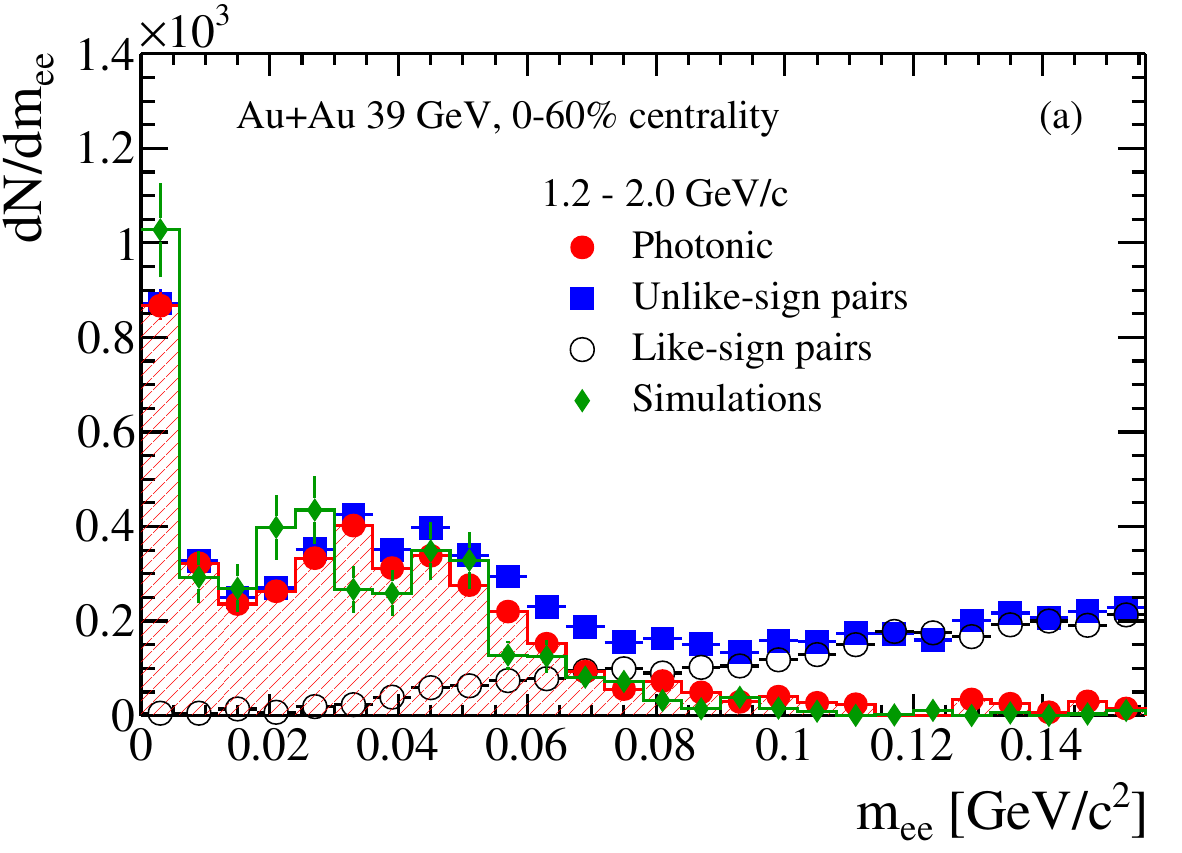} \\
\includegraphics[width=0.45\textwidth]{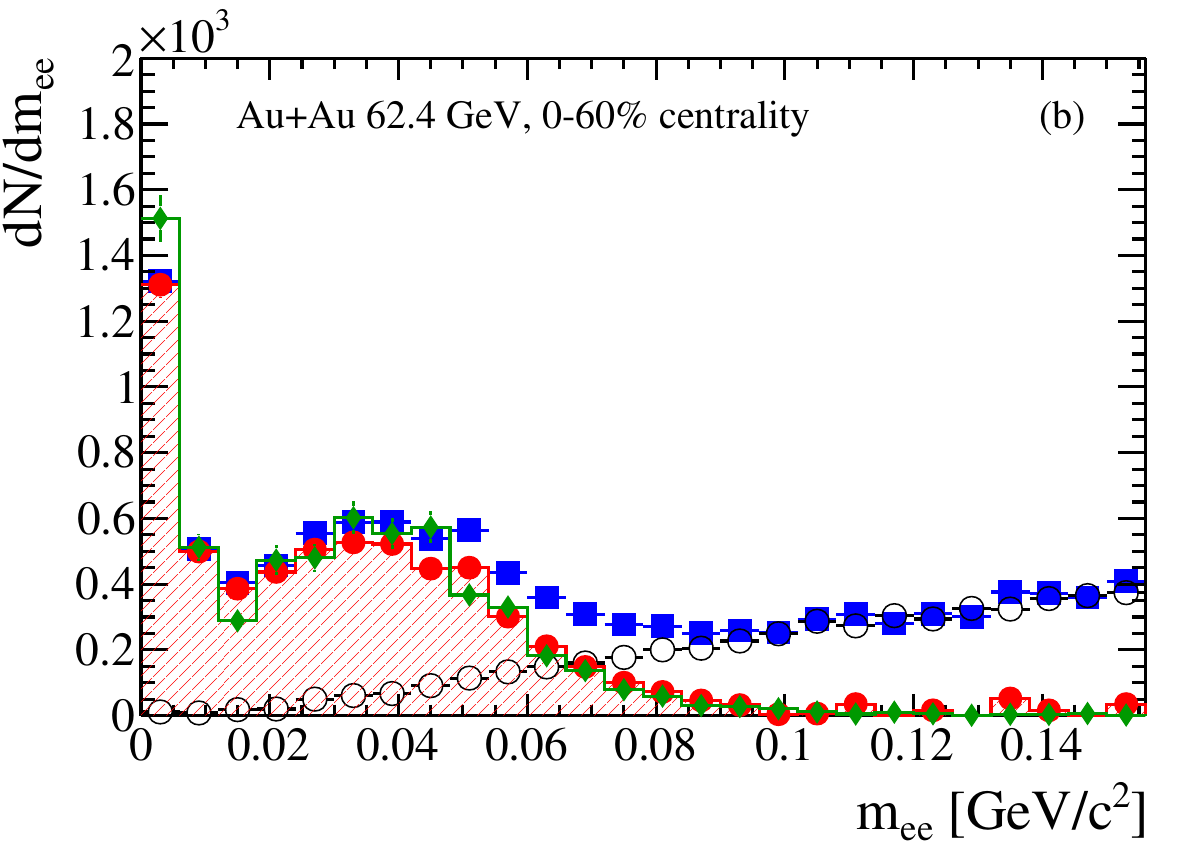} \\
\includegraphics[width=0.45\textwidth]{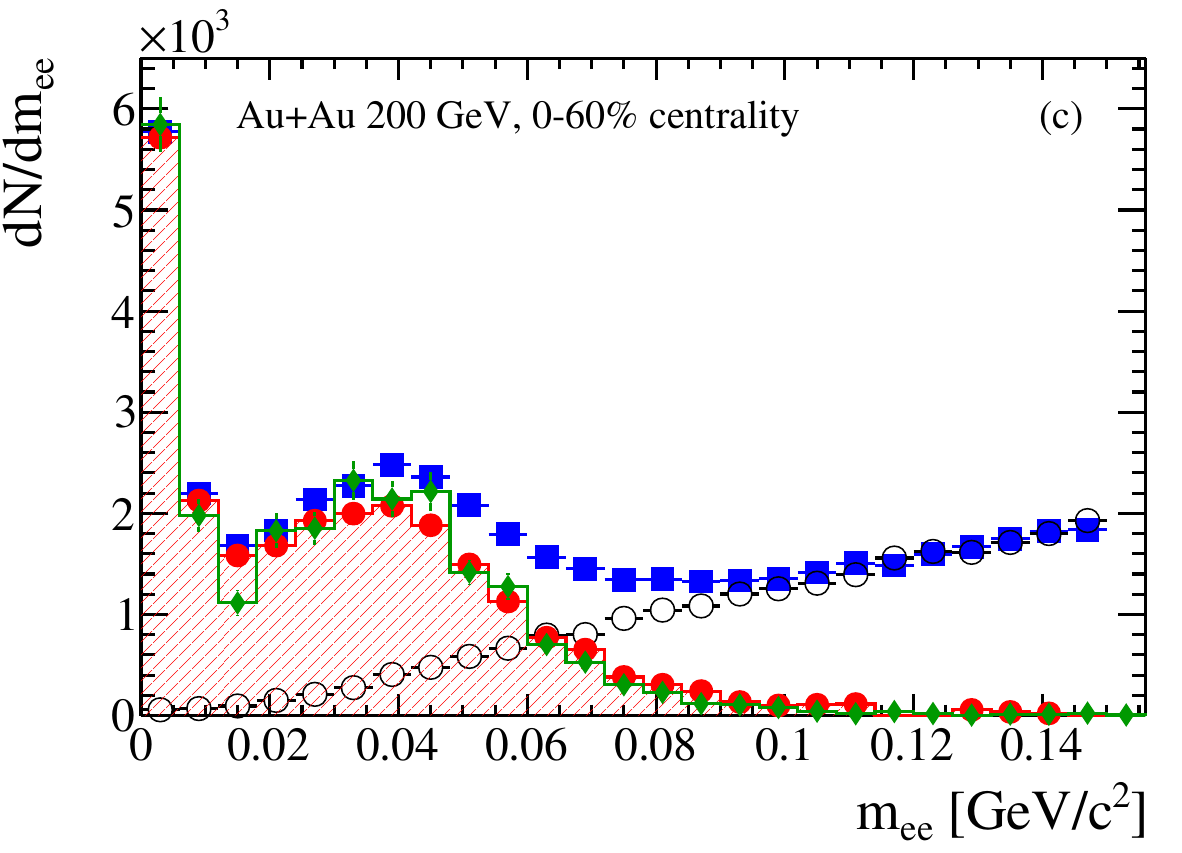} \\
\end{tabular}
\caption{\label{Fig:PhoEleMass}(Color online) Electron pair invariant mass distribution for electrons with $1.2 <\pt<2 \ \gevc$ for the $0-60\%$ most central Au+Au collisions at \sNN{39}~GeV (a), \sNN{62.4}~GeV  (b) and \sNN{200}~GeV (c).} 
\end{center}
\end{figure}

The primary source of physical background for this analysis are so-called photonic electrons. These electrons originate from real photon conversion in the detector material or from Dalitz decay of light mesons (mostly $\pi^0$ and $\eta$). The material thickness relevant for the photon conversion background in STAR in 2010 amounts to $1.05 \%$ of a radiation length. It comes mostly from the beam pipe ($0.29\%$), the inner field cage ($0.45\%$) and a wrap around the beam pipe ($0.17\%$)~\cite{STAR:NPE:pp200GeV}. We identify photonic electrons using a statistical approach, as a signal in the low mass region of the di-electron $m_{e+e-}$ mass spectrum (mass $m_{e+e-}<0.15$~$\rm{\gevc}^2$)~\cite{STAR:NPE:pp200GeV}. Each primary photonic electron candidate is paired with an opposite-sign electron (so-called partner) in an event. We estimate the combinatorial background in this procedure with the like-sign technique, by taking all possible $e^+e^+$ and $e^-e^-$ pairs in an event and adding these two distributions together. Figure \ref{Fig:PhoEleMass} shows examples of  $m_{e+e-}$ distributions for minimum-bias Au+Au collisions at \sNN{39}, 62.4 and 200 GeV. The photonic electron yield is calculated by $\nPho = (N^{\rm UL} - N^{\rm LS})/\effPho$, where $N^{\rm UL}$ and $N^{\rm LS}$ are the numbers of unlike-sign and like-sign electron pairs respectively, and $\effPho$ is the partner finding efficiency (also called the photonic electron tagging efficiency). This method assumes that there is no contribution from correlated hadron pairs at the low invariant mass range. It has been demonstrated~\cite{Kikola:2015qpa} that the effect of correlated hadron pairs on the photonic electron yield calculations is negligible with the invariant mass cut and purity level in our measurement. The $\effPho$ was determined from full GEANT simulations of the STAR detector, which include $\pi^0$ and $\eta$ Dalitz decays and $\gamma$ conversions in the detector material. We use the measured pion ($\pi^{\pm}$ and $\pi^0$) and direct photon $\pt$ spectra as an input in these simulations. Figure \ref{Fig:PurityPheRecoEff} (b) shows $\effPho$ as a function of $\pt$; it varies from 15\% at 0.5~\gevc\ to 60\% at 7 \gevc. 

The ``raw" number of electrons from heavy-flavor decays, $\nHFE$, is given by $\nHFE = pN_I - \nPho$, where $N_I$ is the inclusive electron candidate yield and $p$ is the purity. Besides photonic electrons, other sources of background in this analysis are weak kaon decay ($K^{\pm} \rightarrow e^{\pm}\nu\pi^{0}$ and $K^{0}_{L} \rightarrow e^{\pm}\nu\pi^{\mp}$), called $\Ke$, Drell-Yan, quarkonia and other vector mesons~\cite{STAR:NPE:pp200GeV}. $\Ke$ is the largest source of that secondary background at low $\pt$ ($\pt< 1 \ \gevc$), and we subtract it from our electron sample, as described later in this section. The contribution from $J/\psi \rightarrow e^+e^-$ decays is less than 1\% at $\pt<2~ \gevc$ and increases with $\pt$ to 20\% at $\pt \approx 7 \, \gevc$. This contribution is expected to be approximately energy independent because $D \rightarrow e$ and $J/\psi \rightarrow e^+e^-$ yields depend on the total cross section for charm production in a similar way. The Drell-Yan production and $\Upsilon$ decays play a negligible role with a less than 1\% effect.

The vector meson $(\omega \rightarrow e^+e^-, \pi^0e^+e^- , \eta' \rightarrow \gamma e^+e^-, \phi \rightarrow e^+e^-, \rho \rightarrow e^+e^-)$ contribution changes with the energy since the charm cross section decreases faster with decreasing $\sqrt{s}$ than the production of light mesons. We calculate that $\omega, \eta', \phi, \rho$ feed-down contributes 5-10\% of \HFE\ in minimum bias Au+Au collisions at \sNN{200}~GeV, approximately independent of $\pt$. 
At lower energies, the vector meson contribution is estimated to be $\sim 5\%$ at $\pt<0.5 \, \gevc$, increasing to $\sim 15\%$ at 62.4 GeV/c and $\sim 20\%$ at 39 GeV for $0.5 < \pt < 2 \, \gevc$.

\begin{figure}[htp]
\begin{center}
\includegraphics[width=0.45\textwidth]{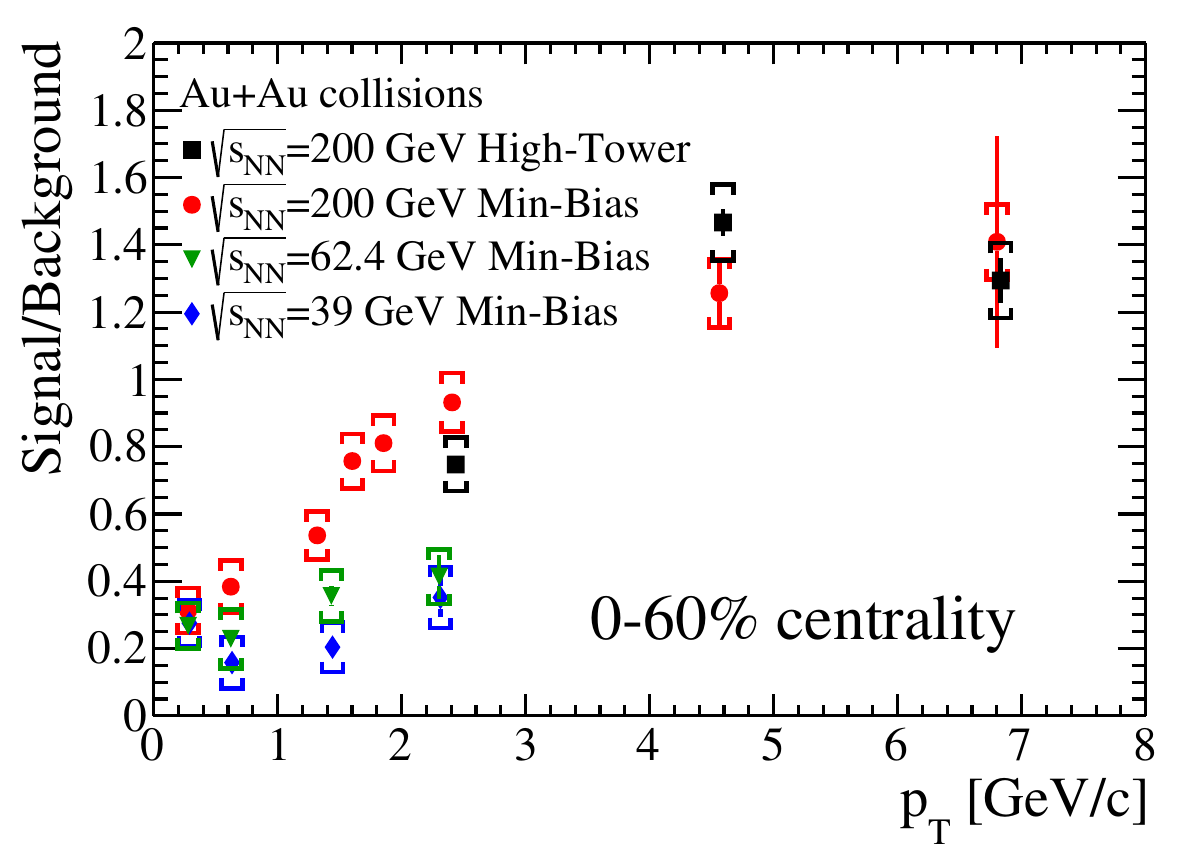} 
\caption{\label{Fig:SignalBackgroundRatio} (Color online) Signal-to-background ratio for electrons from heavy-flavor hadron decays in Au+Au collisions at $\sqrt{s_{\rm NN}} = $ 200, 62.4 and 39 GeV in events with minimum-bias (``Min-Bias'') and high tower (``High-Tower'') triggers. The error bars represent the statistical uncertainty, and the brackets represent the systematic uncertainties. See text for details. 
} 
\end{center}
\end{figure}

Figure \ref{Fig:SignalBackgroundRatio} shows the ratio of the \HFE\ electron signal (with $\Ke$ background subtracted) to the photonic electron background for Au+Au collisions at 200, 62.4 and 39 GeV. At 200 GeV, this ratio varies from 0.3 at low $\pt$ to 1.4 at $\pt$ above 5 \gevc. Overall, this ratio is lower at 62.4 and 39 GeV compared to 200 GeV because the cross section for heavy quark production decreases faster with decreasing colliding energy than does the cross section for the photonic electron background.

Elliptic flow is defined as the second harmonic ($v_2$) in the Fourier expansion of the particle azimuthal anisotropic 
distribution with respect to the reaction plane, $\psiRP$~\cite{Voloshin:1994mz}:
\begin{equation}
%\frac{{\rm d}^2 N}{{\rm d} p_T {\rm d}\phi} \propto 1 + \sum_{n=1}^{\infty} 2v_n (p_T) \cos(n(\phi-\psiRP)) \, ,
\frac{{d}^2 N}{{d} p_T {d}\phi} \propto 1 + \sum_{n=1}^{\infty} 2v_n (p_T) \cos(n(\phi-\psiRP)) \, ,
\end{equation}
where $\phi$ and $p_T$ represent the azimuthal angle and the transverse momentum of the particle, respectively.
The reaction plane is defined with the impact parameter and the beam momenta. In practice, the estimated reaction plane is called the event plane.

To determine the elliptic flow of electrons from heavy-flavor hadron decays, $\vHFE$, we first measure the inclusive electron $v_2^I$, the photonic electron $\vPho$ and the hadron azimuthal anisotropy $v_2^H$ and their yields. Then the $\vHFE$ is given by 
\begin{equation}
\vHFE = \frac{N_I v_2^I - \nPho\vPho - N_H v_2^H}{\nHFE}
\end{equation} 
where $N_H = (1-p)N_I$ is the hadron contamination. $v_2^H$ is calculated as the sum of $v_2$ for different particle species~\cite{STAR:pi:v2:200GeV,STAR:v2:BES,STAR:v2:BES:PRL} weighted by their yields in the inclusive electron sample. These yields are estimated based on the purity studies. The elliptic flow of these components (inclusive and photonic electrons and hadrons) can be measured using any method (for instance $\vn{2}$, $\vn{4}$ or $\vn{EP}$). 

In the $\vn{2}$ and $\vn{4}$ analyses, we obtain $v_2^I$ and $v_2^H$ directly from the data. The inclusive electron $\vn{2}$ and $\vn{4}$ are calculated using the direct cumulant method \cite{Bilandzic:2010jr}: for $\vn{2}$ we correlate an electron with a single hadron, while one electron is correlated with three hadrons for $\vn{4}$.  To optimize the procedure, $\vn{2}$ and $\vn{4}$ of the \HFE\ are calculated with respect to the so-called reference flow~\cite{Bilandzic:2010jr}. 
The reference flow is $v_2$ averaged over some phase space that serves as a reference for $\pt$-differential studies of particles of interest (\HFE\ in this case). We calculate the reference flow using tracks with $0.2 < \pt <2$ \gevc\ within $|\eta| < 1$, excluding tracks with $|\nse| < 3$ to avoid self-correlations. The results are corrected for non-uniform azimuthal detector acceptance by applying the procedure described in Ref.~\cite{Bilandzic:2010jr}.
$\vPho$ is given by GEANT simulations of electrons from $\gamma$ conversions and $\pi^0$ and $\eta$ Dalitz decays, where the measured parent $v_2(\pt)$ and $\pt$ spectra are required as an input. Direct photon $v_2$ values and $\pt$ spectra at 200 GeV are taken from Refs.  \cite{Phenix:LowPt:DirectPhoton,Phenix:HighPt:DirectPhoton,Phenix:DirectPhoton:v2}. For Au+Au collisions at 62.4 and 39 GeV, there are no published direct photon data available; therefore, we use results for \pp\ and assume binary scaling of the direct photon yield. We use next-to-leading-order pQCD calculations for \pp\ at 62.4 GeV~\cite{Phenix:DirPhotonAuAu62GeV,Gordon:1993qc} and  E706 data for 39 GeV~\cite{E706:HighPt:DirectPhoton}. We use the $v_2(\pt)$ ($\vn{2}$ and $\vn{EP}$) and $\pt$ spectra for neutral and charged pions measured by STAR and PHENIX as input for the simulation~\cite{STAR:pi:pTspectra:200GeV,STAR:pi:pTpspectra:62GeV,STAR:pi:v2:200GeV,PHENIX:pi:pTpspectra:200GeV,PHENIX:pi:pTpspectra:62:39GeV,PHENIX:pi:v2:200GeV}. 
The input distributions are parametrized in the simulation: pion spectra are fitted with a power law function $f(\pt) = A(e^{-B\pt - C\pt^2} + \pt/D)^{-n}$, where $A$, $B$, $C$, $D$ and $n$ are fit parameters and we assume $m_T$ scaling for $\eta$. For the direct gamma spectrum, we employ a power law plus exponential fit. The $v_2$ data are parametrized with a $4^{th}$ order polynomial.

In the event-plane analysis, we reconstruct an event plane using tracks with $0.15 < \pt <1.5$ \gevc\ and $|\eta| < 1$ in order to reduce the effect of jets on the event plane estimation. We exclude tracks with $|\nse| < 3$ to avoid possible self-correlations between the particle of interest (the electron) and tracks used in the event plane reconstruction. The results are corrected for non-uniform detector acceptance using $\phi$ weighting and event-by-event shifting of the planes, which is needed to make the final distribution of the event planes isotropic~\cite{Poskanzer:1998yz}. We obtain $\vHFE\{\rm EP\}$ directly from the data: we measure the \HFE\ production differentially at all azimuthal angles with respect to the event plane and fit the distribution with ${d}N/{d}\Delta \phi = A \times [1 + 2 v_2^{\rm observed} \cos (2 \Delta \phi)]$, where $\Delta \phi \equiv \phi - \psiEP$ is the electron azimuthal angle $\phi$ measured with respect to the event plane $\psiEP$, reconstructed event by event. The final $\vHFE\{\rm EP\}$ is calculated by correcting $v_2^{\rm observed}$ with the so-called event plane resolution $R$: $\vHFE\{\rm EP\} = v_2^{\rm observed}/{\it R}$. The event plane resolution is estimated from the correlation of the planes of independent sub-events~\cite{Poskanzer:1998yz} and it is on the level of 0.7 for 0-60\% central events.

\begin{figure*}[htp]
\begin{center}
\includegraphics[width=1.0\textwidth]{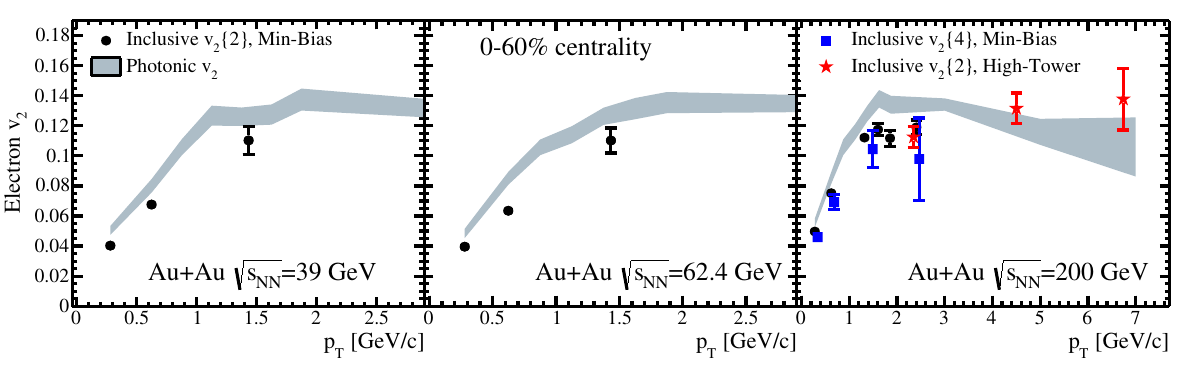} 
\caption{\label{Fig:IncPhoEleV2} (Color online) Inclusive and photonic electron $\vn{2}$ and  $\vn{4}$ at $\sqrt{s_{\rm NN}} = $ 200, 62.4 and 39 GeV. The error bars on the inclusive electron $v_2$ represent the statistical uncertainty. See text for details.} 
\end{center}
\end{figure*}

 The $\Ke$ contribution is estimated using a full GEANT simulation of the STAR detector for both $K^0_L$ and charged kaons. We use the $K_S^0$ $\pt$ spectra measured by STAR~\cite{Aggarwal:2010ig,Agakishiev:2011ar,XiaopingZhang:Sqm2015} as an input in these simulations. The efficiency for $\Ke$ reconstruction is very low at low $\pt$ due to a DCA cut applied in the analysis:  2\% at $\pt=0.5$~\gevc\ and 5\% at $\pt=1$~\gevc. We compared the $\Ke$ background to the expected heavy-flavor decay electron yield taking into account the single electron reconstruction efficiency and acceptance. In the case of Au+Au collisions at 200 GeV, we use the \HFE\ spectra measured by PHENIX \cite{Adare:2010de} as an input. For Au+Au collisions at 39 and 62.4 GeV, the \HFE\ $\pt$ spectrum for low $\pt$ is not available and we use a perturbative QCD prediction for \HFE\ production~\cite{FONLL:RamonaVogt} scaled by the number of binary collisions. The \HFE\ measurements in \pp\ at \sNN{200} GeV are consistent with the upper limit of the pQCD calculation; therefore, we use the upper limit on the predictions as an estimate of \HFE\ yield at lower energies. The $\Ke$ electron background is small at 200 GeV and it decreases with increasing $\pt$: we estimate it to be 8\% for $\pt<1$~\gevc\ and less than 2\% for $\pt>3$~\gevc. However, the heavy quark production cross section decreases faster with decreasing energy than does the cross section for strangeness production. Thus the relative $\Ke$ electron background is larger at 39 and 62.4~GeV than at the top RHIC energy: it amounts to  $\approx 30\%$ for $\pt<0.5$~\gevc\ and  $\approx 10\%$ for $0.5<\pt<3$~\gevc\ at 62.4 GeV. It is even higher at 39~GeV: $\approx 50\%$ for $\pt<0.5$~\gevc\ and $\approx 20\%$ for $0.5<\pt<3$~\gevc. We calculate the $\Ke$ $v_2$ using a GEANT simulation of the STAR detector taking as input the kaon $\pt$ spectrum~\cite{Aggarwal:2010ig,Agakishiev:2011ar,XiaopingZhang:Sqm2015} and $v_2$~\cite{Abelev:2008ae,STAR:v2:BES} measured by STAR. The expected $\Ke$ $\pt$ spectrum and $v_2$ are then subtracted from the measured electron yield and $v_2$. 
 
 There are three dominant sources of systematic uncertainties in this analysis: the photonic electron tagging efficiency, the purity and the input parameters to the photonic electron $v_2$ simulation. We estimated the systematic uncertainty on $\effPho$ by varying the contribution of direct photons to the photonic electron yield (we consider two cases: a negligible direct photon yield or a contribution two times larger than the default), by comparing the  partner finding efficiency in the simulations and the data and by varying the input pion spectra within their statistical and systematic uncertainties. The uncertainties on the input spectra are studied with a Monte Carlo approach. We randomly shift the data points by their combined uncertainties (statistical and systematic) assuming these uncertainties have Gaussian distributions and that $\pt$-bin to $\pt$-bin correlations between systematic uncertainties are insignificant. Then we re-fit the input spectra and we use the fit results as an input in the $\effPho$ calculation. Such a procedure is repeated many times to obtain the $\effPho$ distribution for a given $\pt$ bin. The standard deviation of this distribution for a given $\pt$ is taken as an estimated of systematic uncertainty owing to the precision of input spectra. The partner tagging efficiency is estimated using data in the following way. We assume that efficiencies for different cuts for a partner (number of TPC points on the track, distance of closest approach between photonic electron candidate and a partner, ratio of number of points to the maximum possible) are independent of each other. The efficiency for a given cut is calculated as a ratio of the number of partner tracks that passed a given cut to the number without that condition. Then the photonic electron tagging efficiency is a product of the efficiencies of the different cuts. This approach does not rely on the details of the simulations of photonic electron sources or the STAR detector, but it neglects possible correlations between efficiencies. The relative uncertainty owing to the difference of $\effPho$  in the simulation vs data is less than 6\% and we assign 6\% as a conservative estimate of this uncertainty.  We found that the direct photon contribution and the difference in the value of $\effPho$ obtained from simulations and real data dominate the systematic uncertainty. The overall systematic uncertainty on $\effPho$ is $\pm 7\%$ at 200 GeV, $\pm 8\%$ at 62.4 GeV and $\pm 10\%$ at 39 GeV.  The systematic uncertainty on the purity is estimated by varying the constraints in a multi-Gaussian fit and by changing the fit model for kaons and protons: we used $\nse$ distributions obtained directly from the data using ToF with strict mass cuts instead of Gaussian functions. These uncertainties vary strongly with $\pt$; Fig.~\ref{Fig:PurityPheRecoEff}(a) shows the purity with the combined systematic and statistical uncertainties. The uncertainty on the photonic electron $v_2$ and the $\Ke$ $v_2$ is evaluated by varying the input $\pt$ and $v_2$ spectra within their statistical and systematic uncertainties (employing the same Monte Carlo approach as used for $\effPho$) and varying the relative contributions of the simulation components for the photonic electron $v_2$. The overall uncertainty on the photonic electron $v_2$ is 6\% for $\pt<5$~\gevc. However, at high $\pt$ in Au+Au collisions at $\sqrt{s_{\rm NN}} = $ 200 GeV it increases with $\pt$ to 20\% at $\pt = 7$~\gevc. The uncertainty on the $\Ke$ $v_2$ is $15-20\%$. We estimate the systematic uncertainty on the $\Ke/$\HFE\ ratio by varying the input \HFE\ distribution. At 200 GeV, we vary the input spectra within statistical and systematic uncertainties; at 39 and 62.4 GeV, we use the central value of pQCD predictions as an estimate of the lower limit on the \HFE\ production. Table~\ref{Tab:syst:error} summarizes the uncertainties of various elements of the measurement.

 \begin{table*}[htp]
 %\begin{ruledtabular}
 \begin{tabular}{lccc}
 \hline \hline 
Uncertainties on various elements of the analysis & \multicolumn{3}{c}{Relative uncertainty}  \\

 & \sNN{200}~GeV &  \sNN{62.4}~GeV &  \sNN{39}~GeV  \\
 \hline 
Purity & $1-65\%$ & $1-44\%$ & $1-19\%$ \\
$\effPho$ & $7\%$ & $8\%$ & $10\%$ \\
~~~~-- Direct photon yield & $0.5-6\%$ & $0.5-4\%$ & $0.5-6\%$ \\
~~~~-- Partner finding efficiency in the simulation vs data & $6\%$ & $6\%$ & $6\%$ \\
~~~~-- Input $\pi^0$ and $\eta$ $\pt$ spectrum & $<1\%$ & $<1\%$ & $<1\%$ \\
~~~~-- Statistical uncertainty & $2\%$ & $4\%$ & $5\%$ \\
Photonic electron $v_2$ & $6-20\%$ & $6\%$ & $6\%$ \\
$\Ke$ contribution to \HFE & $1-3\%$ & $1-3\%$ & $1-5\%$ \\
$\Ke$ electron $v_2$ & $15-20\%$ & $15-20\%$ & $20\%$ \\
 \hline \hline 
 \end{tabular}
 %\end{ruledtabular}
 \caption{\label{Tab:syst:error} Main sources of systematic uncertainties of the various elements of the analysis. Most of the uncertainties are $\pt$ dependent.}
 \end{table*}

\section{\label{sec:results}Results}

\begin{figure}[htp]
\begin{center}
\begin{tabular}{c}
\includegraphics[width=0.45\textwidth]{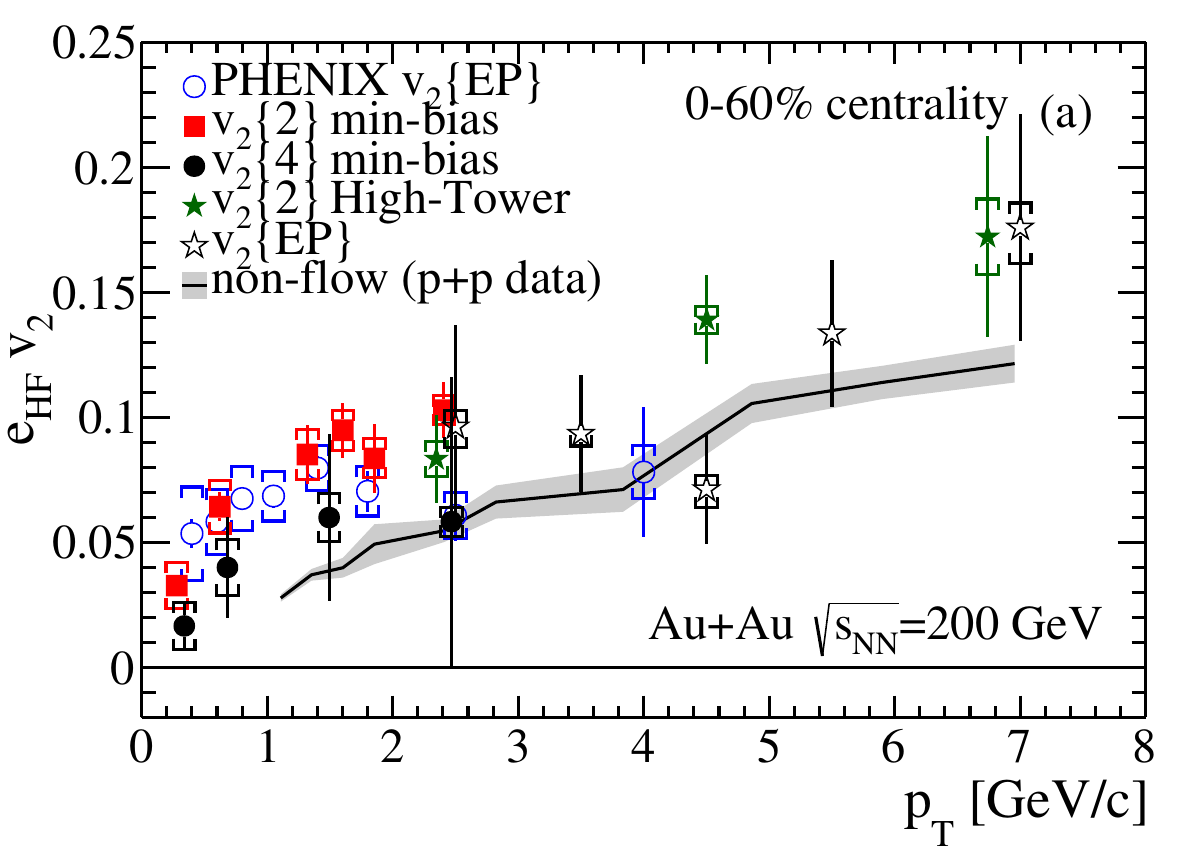} \\
\includegraphics[width=0.45\textwidth]{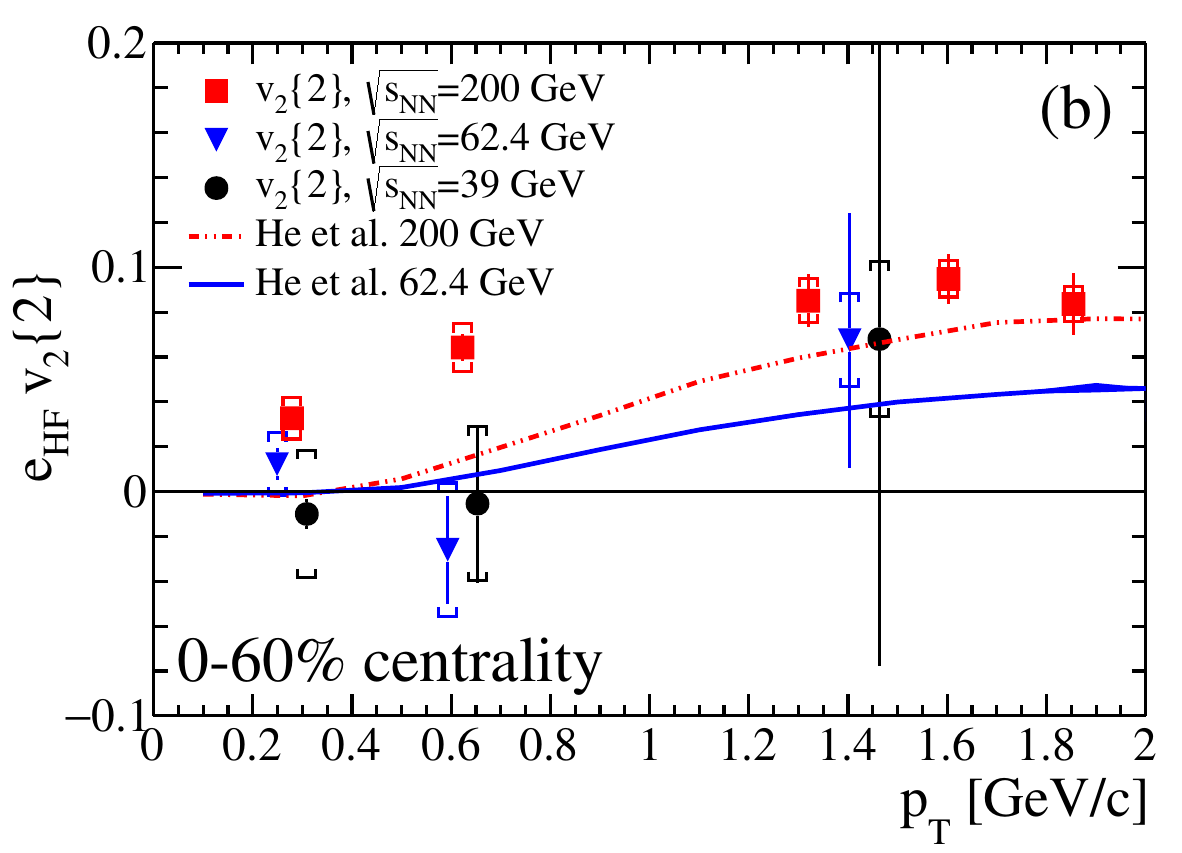} 
\end{tabular}
\caption{\label{Fig:HFEV2} (Color online)(a) Elliptic flow $v_2$ of electrons from heavy-flavor hadron decays at $\sqrt{s_{\rm NN}} = $ 200 GeV compared to PHENIX measurements~\cite{Adare:2010de}. (b) \HFE\ $\vn{2}$ at 200 and 62.4 and 39 GeV. The error bars represent the statistical uncertainty, and the brackets represent the systematic uncertainties. Non-flow in (a) was estimated based on \HFE-hadron correlations~\cite{Aggarwal:2010xp} for $\pt>2.5$~\gevc\ and PYTHIA for $\pt<2.5$~\gevc. The band includes the combined systematic and statistical uncertainties. The curves in (b) show TMatrix model calculations for $\sqrt{s_{\rm NN}} = 62.4$~GeV~\cite{v2:62GeV:He} and 200 GeV~\cite{He:2011qa}. } 
\end{center}
\end{figure}

\begin{figure}[htp]
\begin{center}
\includegraphics[width=0.45\textwidth]{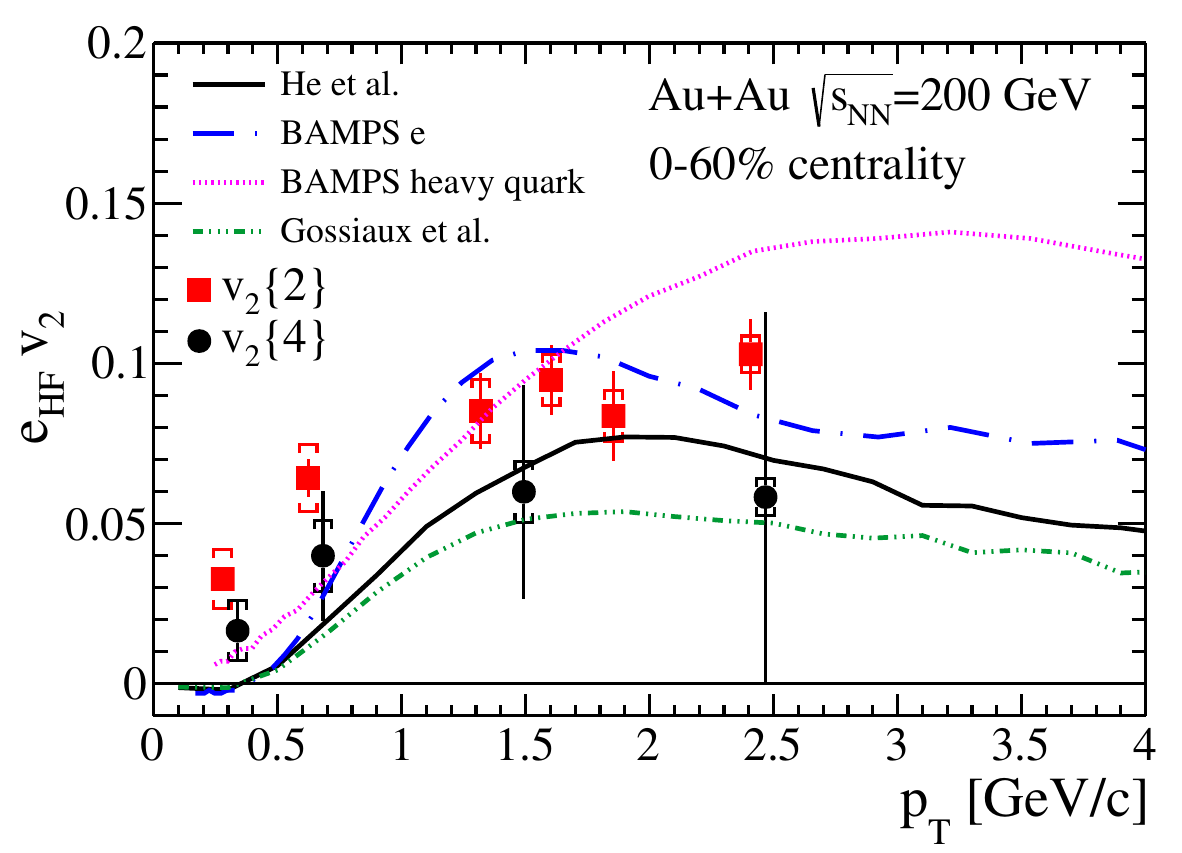} 
\caption{\label{Fig:HFEV2WithModels} (Color online) The \HFE\ elliptic flow $\vn{2}$ and  $\vn{4}$ at $\sqrt{s_{\rm NN}} = $ 200 GeV (min-bias) from Fig. \ref{Fig:HFEV2}(a) compared to model calculations.} 
\end{center}
\end{figure}

Figure \ref{Fig:IncPhoEleV2} shows the inclusive and photonic electron $\vn{2}$ and $\vn{4}$ for the 0-60\% most central Au+Au collisions at 200, 62.4 and 39 GeV. The photonic electron $v_2$ is larger than the inclusive electron $v_2$ at low and intermediate $\pt$ ($\pt< 4$~\gevc), which indicates that the \HFE\ $v_2$ has to be smaller than $v_2^I$. Figure \ref{Fig:HFEV2} shows the \HFE\ elliptic flow $v_2$ at $\sqrt{s_{\rm NN}} = $ 200 GeV (a), and 62.4 and 39 GeV (b). We observe positive $\vn{2}$ and $\vn{4}$ for $\pt>0.5$~\gevc\ at 200~GeV. At high $\pt$, the $\vn{2}$ and $\vn{EP}$ results are consistent with each other, as expected. There is a hint of an increase of $v_2$ with $\pt$ for $\pt>4$~\gevc, which is probably an effect of jet-like correlations. We estimate the strength of these correlations for $\pt>2.5$~\gevc\ using \HFE--hadron correlations in \pp\ at $\sqrt{s}=$~200 GeV~\cite{Aggarwal:2010xp}; the non-flow correlations in \pp\ are scaled by the hadron multiplicity in Au+Au collisions, similarly to Ref.~\cite{STAR:highpT:v2}. If we assume that the non-flow correlations in \pp\ are similar to those in Au+Au collisions, then the non-flow in Au+Au reactions can be estimated by 
\begin{equation}
v_2^{\rm non-flow} = \frac{ \langle\langle2'\rangle\rangle^{pp}}{\vn{2}^{\rm Ref}}\frac{\langle N_h^{pp} \rangle}{\langle N_h^{\rm AA} \rangle}\, , 
\end{equation}
where $\langle\langle2'\rangle\rangle^{pp}$ is the average two-particle correlation of \HFE\ and hadrons in \pp, $\langle N_h^{pp} \rangle$ and $\langle N_h^{\rm AA} \rangle$ are the average number of hadrons in \pp\ and Au+Au collisions, respectively, and $\vn{2}^{\rm Ref}$ is the reference $v_2$ in Au+Au collisions. The jet-like correlation may be considerably modified in the QGP, therefore this procedure likely gives a conservative estimate of the non-flow. 

We found that PYTHIA simulations, with the trigger and single track reconstruction efficiencies included, reproduce well the $v_2^{\rm non-flow}$ obtained with \pp\ data at 200 GeV. Thus we use PYTHIA to estimate the $v_2^{\rm non-flow}$ for $\pt < 2.5$~\gevc. The black solid line in Fig. \ref{Fig:HFEV2} (a) shows the jet-like correlations expected in Au+Au collisions, with the gray band representing the statistical uncertainties combined with the systematic uncertainties due to electron identification and photonic electron rejection~\cite{Aggarwal:2010xp}. Those correlations can explain the rise of $\vn{2}$ and $\vn{EP}$ with $\pt$; more than 60\% of the $v_2$ signal at high $\pt$ could be explained by the central value of non-flow (black solid line in Fig. \ref{Fig:HFEV2} (a)). This indicates that ``conventional'' jet correlations (i.e. correlations unrelated to the reaction plane) are likely to dominate $v_2$ for $\pt>4$ \gevc. We did not estimate the jet-like correlation at 39 and 62.4~GeV because the \HFE--hadron correlation data are not available at those energies.     

STAR data are compared to PHENIX measurements for $|\eta|<0.35$ in Fig.~\ref{Fig:HFEV2}(a). PHENIX used beam-beam counters~(BBCs) with a pseudorapidity coverage of $3.0<|\eta|<3.9$ to measure the event plane. A large pseudorapidity gap between the BBCs and the detector used for electron identification is expected to reduce the effect of jet-like correlations and resonance decays on the $v_2$ measurement. PHENIX data are consistent with STAR results in the $\pt$ range where they overlap ($\pt\leq4$~\gevc). The ALICE collaboration also measured the heavy-flavor decay electron $v_2$ in Pb+Pb collisions at $\sqrt{s_{NN}} =2.76$~TeV~\cite{ALICE:HF:v2} using an event plane method and the observed elliptic flow at low and intermediate $\pt$ $(\pt<5 \, \gevc)$ is similar to that at RHIC. At higher $\pt$, the $v_2$ in Pb+Pb collisions decreases with increasing transverse momenta, contrary to our results. The ALICE collaboration uses an event plane method with a rapidity gap of $|\Delta \eta|>0.9$ which reduces non-flow correlations. Thus, the high-$\pt$ trend observed by STAR suggests a contribution of jet-like correlations to the measured $v_2$.

At 39 and 62.4 GeV, $\vn{2}$ is consistent with zero up to $\pt=1.6$~\gevc\ (see Fig.~\ref{Fig:HFEV2}(b)). We further check if the $v_2$ values observed for the two lower energies deviate significantly from the trend seen at the top RHIC energy. We quantify the difference using the $\chi^2$ test to verify the null hypothesis that the $\vn{2}$ at 200 GeV is consistent with those at 62.4 and 39 GeV for $\pt<1$ \gevc. We define the test-statistic as
\begin{equation}
%\chi^2 = \sum\limits_{\pt < 1 \,{\rm \gevc}} \frac{\left(v_2^{\rm 200\, GeV} - v_2^{\rm 39 (62.4) \, GeV}\right)^2}{\sigma^{2}_{\rm 200\, GeV} + \sigma^{2}_{\rm 39 (62.4) \, GeV}}
\chi^2 = \sum\limits_{\pt < 1 \, {\rm GeV/}c} \frac{\left(v_2^{\rm 200 \, GeV} - v_2^{\rm lower}\right)^2}{\sigma^{2}_{\rm 200\, GeV} + \sigma^{2}_{\rm lower}}
\end{equation} 
where $v_2^{\rm lower}$ and $\sigma_{\rm lower}$ denote $v_2$ and $\sigma$ for lower energies, $\sigma = \sqrt{\sigma_{\rm stat.}^2 + \sigma_{\rm syst.}^2}$, the number of degrees of freedom, NDF, is 2, and we assumed that these two samples are independent of one another and the uncertainties have normal distributions. The $\chi^2$/NDF value for a consistency between 200~GeV and 62.4~GeV is 6.3/2 which corresponds to a probability $p = 0.043$ of observing a $\chi^2$ that exceeds the current measured $\chi^2$ by chance. For the comparison between 200 and 39~GeV, $\chi^2/{\rm NDF} = 3.82/2$ which corresponds to $p = 0.148$. 
PHENIX reported that the measured $v_2$ of heavy flavor decay electrons in Au+Au collisions at \sNN{62.4}~GeV is positive when averaged across $\pt$ between 1.3 and 2.5 \gevc ~\cite{PHENIX:HF:v2:62GeV}. However, the PHENIX $v_2$ result is less than $1.5\sigma$ away from zero when systematic and statistical uncertainties are taken into account (Fig. 23 in Ref.~\cite{PHENIX:HF:v2:62GeV}). PHENIX $\vn{EP}$ measurements in Au+Au collisions at \sNN{62.4}~GeV agree with STAR results in the overlapping $\pt$ range within sizable uncertainties. 
 
Contrary to the results for light hadrons, for which a positive $v_2$ is observed and the difference between \sNN{200}~GeV and 39 GeV is small, our measurements in Au+Au collisions at \sNN{62.4}~GeV and 39 GeV indicate that the $v_2$ of electrons from heavy flavor hadrons decays is consistent with zero. Moreover, the $v_2$ for \HFE\ at both \sNN{39} and 62.4 GeV is systematically lower than at \sNN{200}~GeV for $\pt<1 \ \gevc$. 

The observed $v_2$ for \HFE\ is modified with respect to the parent quark $v_2$ due to the decay kinematics of the parent heavy hadron. This effect is shown in Fig.~\ref{Fig:HFEV2WithModels} by the predictions for heavy quark elliptic flow and the resulting electron $v_2$ from the partonic transport model BAMPS (Boltzmann approach to multiparton scatterings)~\cite{Uphoff:2011ad,Uphoff:2012gb}. The \HFE\ production at low transverse momenta is dominated by charm hadron decays~\cite{Aggarwal:2010xp}. 

Although the PYHTIA simulation shows that the correlation between an azimuthal angle of \HFE\ and the parent D-meson decreases with decreasing $\pt$ due to the D-meson decay kinematics, there is still a correlation even at $\pt \sim 0.2 \, \gevc$. Therefore, the observed difference of $v_2$ values may indicate that charm quarks interact less strongly with the surrounding nuclear matter at these two lower energies compared to \sNN{200}~GeV. However, more data are required to draw definitive conclusions.

As discussed before, the \HFE\ $v_2$ is modified with respect to the parent quark $v_2$. Also, the \HFE\ $\pt$ spectrum is shifted towards lower $\pt$ compared to the parent hadron spectra, which makes the interpretation of the \HFE\ data model-dependent. Figure \ref{Fig:HFEV2WithModels} shows the \HFE\ $\vn{2}$ and $\vn{4}$ at 200 GeV compared to a few models of heavy quark interactions with the partonic medium, which are described below. Note that all models here calculate the elliptic flow of \HFE\ and heavy quarks with respect to the reaction plane. The flow fluctuations and non-flow are not included there, therefore the predicted $v_2$ values should be between $\vn{2}$ and $\vn{4}$. Unfortunately, limited statistics do not allow us to quantify this difference in the data -- the measured $\vn{4}$ is consistent with $\vn{2}$ within uncertainties.

In a partonic transport model, BAMPS \cite{Uphoff:2011ad,Uphoff:2012gb} (blue dash-dotted line in Fig.~\ref{Fig:HFEV2WithModels}), heavy quarks lose energy by collisional energy loss with the rest of the medium. To account for radiative energy loss, which is not implemented in this model, the heavy quark scattering cross section is scaled up by a phenomenological factor, K = 3.5. In BAMPS, the hadronization is implemented as fragmentation into $D$ and $B$ mesons using the Peterson function. Thus the observed positive $v_2$ of \HFE\ comes only from the  elliptic flow of charm quarks. Indeed, heavy quarks have a large elliptic flow in this model (dotted line). Note that the Peterson fragmentation is not an appropriate description of hadronization at low $\pt$ and other, more sophisticated mechanisms (for instance, coalescence) should be implemented. Overall, BAMPS describes the $\vn{2}$ data well, but it slightly underestimates  the nuclear modification factor $R_{\rm AA}$ for heavy-flavor decay electrons, reported by PHENIX, at intermediate $\pt$ ($1.5<\pt<4$~\gevc)~\cite{Uphoff:2012gb}. %This is a region where the initial-state parton-$k_T$ broadening (also called the Cronin effect) is the most significant. 
It has been shown in Ref.~\cite{Gossiaux:2008jv} that initial-state parton-$k_T$ broadening (also called the Cronin effect) increases the predicted $R_{\rm AA}$ in a $\pt$ range of 1 - 3 \gevc\ and improves the agreement with the data. However, it has almost no effect at high $\pt$ and thus it is not important for the energy loss studies. 

The dash-dotted green line in Fig.~\ref{Fig:HFEV2WithModels} shows the implementation of radiative and collisional energy loss from Gossiaux et al. \cite{Gossiaux:2008jv,Gossiaux:2010yx,Aichelin:2012ww}. It is a QCD-inspired model with the pQCD description of heavy quark quenching and additional non-perturbative corrections, with the hadronization implemented as coalescence at low $\pt$ and pure fragmentation for high momentum quarks. In this model, there is little contribution from the light quark to the heavy meson $v_2$ and almost all the $D$ or $B$ meson elliptic flow comes from the charm and bottom $v_2$. This model describes the \HFE\ nuclear modification factor at RHIC well. It underpredicts the $\vn{2}$ at intermediate $\pt$, but there is a reasonable agreement with the $\vn{4}$ data. Nevertheless, it predicts a positive \HFE\ $v_2$, which indicates a positive charm quark $v_2$. 

The TMatrix interactions model \cite{vanHees:2007me, He:2011qa} is a non-perturbative approach to heavy quark energy loss. In this framework, the heavy quark interaction with the medium is simulated with relativistic Fokker-Planck-Langevin dynamics for elastic scattering in a strongly coupled QGP (modeled by relativistic hydrodynamics). The model assumes strong coupling between heavy quarks and the bulk medium; hadronization is implemented by combining recombination and fragmentation. In this model, heavy quark resonances are formed in the medium at temperatures up to 1.5 times the critical temperature $T_c$, and scatter off the light quarks in the QGP. The resonant rescattering increases the relaxation rates for charm quarks compared to pQCD scattering of quarks and gluons. This approach also successfully describes the nuclear modification factor and there is a good agreement with the $\vn{4}$ data, although it misses the $\vn{2}$ data points at intermediate $\pt$~(solid black line). The model predicts a moderate difference between $v_2$ in Au+Au collisions at \sNN{200} and 62.4~GeV at low $\pt$ and the calculation for $v_2$ at \sNN{62.4}~GeV\cite{v2:62GeV:He} in Fig.~\ref{Fig:HFEV2}(b) is consistent with our data.

Note that $v_{2}$ should be sensitive to the heavy quark hadronization mechanism. M.~He et al.~\cite{He:2011qa} and P.B.~Gossiaux et al.~\cite{Gossiaux:2008jv,Gossiaux:2010yx,Aichelin:2012ww} use a coalescence approach in the shown $\pt$ range, while in the BAMPS model heavy quarks fragment into mesons. In general, coalescence is expected to give a larger $v_2$ of the mesons due to the contribution of the light quark flow. However, it is shown in~\cite{Greco:2003vf,Adare:2010de} that elliptic flow of light quarks alone cannot account for the observed \HFE\ $v_2$. The data are approximately reproduced if in the model~\cite{Greco:2003vf} charm quarks have an elliptic flow similar to that of light quarks.

The theoretical models discussed here, despite the different mechanisms employed, assume that charm quarks are strongly coupled with the medium and have a positive elliptic flow. All these models qualitatively follow the trend of the data. To further discriminate between models, a simultaneous comparison with other experimental observables (nuclear modification factor, azimuthal correlations) as a function of beam energy is required. 
Moreover, precision measurements of these quantities for charmed and bottom hadrons separately are necessary to further constrain the models and to advance our understanding of the partonic medium properties. Two new STAR detectors, the Heavy Flavor Tracker and the Muon Telescope Detector~\cite{QM2012:STAR:upgrade}, will deliver such data in the next few years.

\section{\label{sec:summary}Summary}

We measured the azimuthal anisotropy $v_2$ of heavy flavor decay electrons over a broad range of energy, starting from the point where the quark gluon plasma state is observed. We report the first measurement of azimuthal anisotropy of electrons from heavy-flavor hadron decays using 2- and 4-particle correlations at $\sqrt{s_{\rm NN}} = $ 200 GeV, and $\vn{2}$ at 62.4 and 39 GeV. \HFE\ $\vn{2}$ and $\vn{4}$ are non-zero at low and intermediate $\pt$  at 200 GeV; more data are needed to quantify the effect of fluctuations and non-flow on the measured elliptic flow. At lower energies, the measured value of $\vn{2}$ is consistent with zero and systematically smaller than those at $\sqrt{s_{\rm NN}} = $~200~GeV for $\pt<1 $~\gevc, although more data are required before one can draw definite conclusions. The difference between \HFE\ $v_2$ observed at $\sqrt{s_{\rm NN}} = $~62.4~GeV and 39~GeV at low traverse momenta and that at $\sqrt{s_{\rm NN}} = $~200~GeV may suggest that charm quarks interact less strongly with the surrounding nuclear matter at these two lower energies compared to \sNN{200}~GeV. However, additional high-precision measurements in a broader $\pt$ range are required to validate this hypothesis.

\section*{Acknowledgements}

We thank the RHIC Operations Group and RCF at BNL, the NERSC Center at LBNL, and the Open Science Grid consortium for providing resources and support. This work was supported in part by the Office of Nuclear Physics within the U.S. DOE Office of Science, the U.S. National Science Foundation, the Ministry of Education and Science of the Russian Federation, National Natural Science Foundation of China, Chinese Academy of Science, the Ministry of Science and Technology of China and the Chinese Ministry of Education, the National Research Foundation of Korea, GA and MSMT of the Czech Republic, Department of Atomic Energy and Department of Science and Technology of the Government of India; the National Science Centre of Poland, National Research Foundation, the Ministry of Science, Education and Sports of the Republic of Croatia, RosAtom of Russia and German Bundesministerium fur Bildung, Wissenschaft, Forschung and Technologie (BMBF) and the Helmholtz Association.

\bibliographystyle{apsrev} 
%\biboptions{sort&compress}
\bibliography{NpeV2}

\end{document}